\documentstyle[twoside,fleqn,espcrc2,epsfig,times]{article}

\newcommand{\eqref}[1]{(\ref{#1})}
\newcommand{\gesim}{\,\raisebox{-.3ex}{$_{\textstyle >}\atop^{\textstyle\sim}$}\,}
\newcommand{\tesla}{\textsc{Tesla}\ }
\newcommand{\gev}{{\rm \ GeV}}
\newcommand{\mev}{{\rm \ MeV}}
\newcommand{\re}{\Re e \,}
\newcommand{\im}{\Im m \,}
\newcommand{\Eslash}{{\not{\!\!E}}}

\newcommand{\sll}{{\tilde{l}}}
\newcommand{\slR}{{\tilde{l}_{\rm R}}}
\newcommand{\slL}{{\tilde{l}_{\rm L}}}

\newcommand{\smuR}{{\tilde{\mu}_{\rm R}}}

\newcommand{\seR}{{\tilde{e}_{\rm R}}}
\newcommand{\seL}{{\tilde{e}_{\rm L}}}
\newcommand{\mseR}{m_{\tilde{e}_{\rm R}}}
\newcommand{\mseL}{m_{\tilde{e}_{\rm L}}}
\newcommand{\cha}{\tilde{\chi}}
\newcommand{\neu}{\tilde{\chi}^0}

\newcommand{\knickpfeil}{\;\raisebox{1.05ex}{$\lfloor$} \!\!\! \to}

\newcommand{\sfR}{{\tilde{f}_{\rm R}}}
\newcommand{\sfL}{{\tilde{f}_{\rm L}}}
\newcommand{\sF}{{\tilde{f}}}
\newcommand{\msf}{m_{\rm \tilde{f}}}
\newcommand{\ff}{f}

\title{\vspace{-2em}
Sfermion Precision Measurements at a Linear Collider\thanks{Based on
talks given by A.F. at
the 31st International Conference on High Energy Physics (ICHEP '02), Amsterdam,
July 2002, and J.K at
the International Workshop on Linear Colliders, Jeju Island, Korea, August 2002
on behalf of the SUSY working groups of the ECFA/DESY and
International Linear Collider Workshops;
to appear in the proceedings.}
\hfill\raisebox{1em}{\mbox{\small DESY--02--176}}}

\author{A.~Freitas\address[desy]{Deutsches Elektronen-Synchrotron DESY, D-22603 Hamburg, Germany}%
	$^{,}$\address{Fermi National Accelerator Laboratory, Batavia, IL
	60510-500, USA}%
	\thanks{afreitas@fnal.gov},
	J.~Kalinowski\address{Insitute of Theoretical Physics, Warsaw
	University, 00681 Warsaw, Poland},
	B.~Ananthanarayan\address[zeuthen]{Deutsches Elektronen-Synchrotron DESY, D-15738
	Zeuthen, Germany},
	A.~Bartl\address[uwien]{Universit\"at Wien, A-1090 Vienna,
	Austria},
	G.~A.~Blair\address{Royal Holloway and Bedford New College, University of
	London, UK},
	C.~Bl\"ochinger\address[wu]{Universit\"at W\"urzburg, D-97074 W\"urzburg,
	Germany},
	E.~Boos\addressmark[desy]$^{,}$\address{Moscow State University, 119899
	Moscow, Russia},
	A.~Brandenburg\addressmark[desy],
	A.~Datta\address[montp]{Universit\'e de Montpellier II, F-34095 Montpellier
	Cedex 5, France},
	A.~Djouadi\addressmark[montp],
	H.~Fraas\addressmark[wu],
	J.~Guasch\address[psi]{Paul-Scherrer-Institute, CH-5232 Villigen PSI,
	Switzerland},
	S.~Hesselbach\addressmark[uwien],
	K.~Hidaka\address{Tokyo Gakugei University, Koganei, Tokyo 184-8501,
	Japan},
	W.~Hollik\address{Max-Planck-Institut f\"ur Physik, D-80805 M\"unchen,
	Germany},
	T.~Kernreiter\addressmark[uwien],
	M.~Maniatis\addressmark[desy],
	A.~v.~Manteuffel\addressmark[desy],
	H.-U.~Martyn\address{RWTH Aachen, D-52056 Aachen, Germany},
	D.~J.~Miller\address{CERN, CH-1211 Gen\`eve 23, Switzerland},
	G.~Moortgat-Pick\addressmark[desy],
	M.~M\"uhlleitner\addressmark[psi],
	U.~Nauenberg\address{University of Colorado, Boulder, CO, USA},
	H.~Nowak\addressmark[zeuthen],
	W.~Porod\address{Universit\"at Z\"urich, CH-8057 Z\"urich, Switzerland},
	J.~Sol\`a\address{Universitat de Barcelona, E-08007 Barcelona, Spain},
	A.~Sopczak\address{Lancaster University, Lancaster LA1 4YW, UK},
	A.~Stahl\addressmark[zeuthen],
	M.~M.~Weber\addressmark[psi],
	P.~M.~Zerwas\addressmark[desy]}

\begin{document}

\begin{abstract}
At future $e^\pm e^-$ linear colliders, the event rates and
clean signals of scalar fermion 
production---in particular for the scalar
leptons---allow very precise measurements of their masses and couplings
and the determination of their quantum numbers. Various methods are proposed
for extracting these parameters from the data at the sfermion thresholds
and in the continuum. At the same time, NLO radiative corrections and
non-zero width effects have been calculated in order to match the
experimental accuracy.  The substantial mixing expected for the third
generation sfermions
opens up additional opportunities. Techniques are presented for
determining potential CP-violating phases and for extracting $\tan\beta$
from the stau sector, in particular at high values.
The consequences of possible large mass differences in the stop and sbottom
system are explored in dedicated analyses.%
\vspace{-2em}%
\end{abstract}

\maketitle

\section{INTRODUCTION}

While the Standard Model of electroweak and strong interactions has been widely
tested and stringently established in various experiments, it nevertheless
exhibits
conceptual drawbacks in the context of Grand Unified Theories and it does not
include gravity.
The introduction of supersymmetry provides a first step to resolve
these problems.
The Minimal Supersymmetric Standard Model (MSSM) %is considered, which
comprises the minimal particle content necessary for extending the Standard
Model into a supersymmetric theory.
It is clear that supersymmetry is broken in nature, but the
construction of a viable breaking mechanism remains a difficult
issue. 
In most scenarios the breaking mechanism resides in an experimentally
inaccessible ``hidden sector'' and is transmitted to the visible world,
thus generating explicit soft-breaking terms in the effective Lagrangian.

If supersymmetric particles are detected in the future, exploring the
underlying theory will include two central tasks. 
On the one hand, to establish
symmetry scheme experimentally, it is necessary to accurately test the
predicted symmetry
relations, in particular the quantum numbers and couplings of the superpartners
must coincide with the corresponding parameters of the Standard Model. On the other
hand, the pattern of supersymmetry breaking needs to be explored.
Because of the large number of soft-breaking parameters in the MSSM, this will
be an enormous experimental effort.
In this context, it is important to determine the soft-breaking parameters with high
precision in order to reconstruct the underlying breaking mechanism,
which eventually involves extrapolations to high energy scales.
A high-energy $e^+e^-$ linear collider \cite{LC} is particularly suited for
precision measurements of supersymmetric particle properties up to the
per-mille level, due to the well-defined
initial state, the high luminosity and the polarization for both beams
\cite{gudi}.
In order to match the experimental accuracy, sufficiently precise and reliable
theoretical predictions for the production cross-sections and decays of the
superpartners are required.

This report is focused on recent developments in precision analyses of scalar
fermions---sleptons and squarks---at a linear
collider. In particular the determination of their quantum numbers, couplings
and masses will be discussed, the latter being the essential building block for
the exploration of the soft-breaking terms. Moreover, due to the large Higgs
Yukawa coupling of the third generation sfermions,
this sector can yield valuable information about the Higgs parameters $\mu$ and
$\tan\beta$ as well as the trilinear scalar couplings $A_{\rm f}$,
including potential CP-violating phases.

\section{DECAY AND PRODUCTION MODES} \label{sec:phen}

\begin{figure}[tb]
\psfig{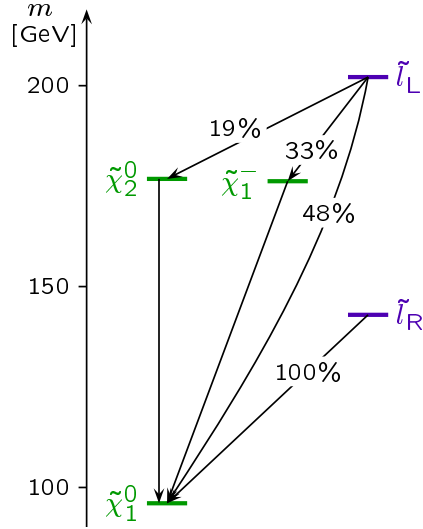}
\vspace{-2em}
\caption{Masses and branching ratios of sleptons of the first two generations in
the SPS1 scenario \cite{sps}. The total widths are $\Gamma_{\!\slR} = 210 \mev$
and $\Gamma_{\!\slL} = 250 \mev$.}
\label{fig:sp}
%\vspace{-1em}
\end{figure}

The phenomenology of scalar leptons is characterized by clean final state
signatures with only a few visible leptons and missing energy. The scalar
partners $\sfL$, $\sfR$ of the left- and
right-handed states of the fermion $f$ have
different decay modes which can be used to discriminate between the two
chirality states:
\begin{equation}
\begin{array}{l}
\sfR^- \to \ff^- \, \neu_1, \quad \sfL^- \to \ff^- \, \neu_2 \\
\phantom{\sfR^- \to \ff^- \, \neu_1, \quad \sfL^- \to \ff^-\,}
  \knickpfeil \tau^+ \tau^- \neu_1\\
\phantom{\sfR^- \to \ff^- \, \neu_1, \quad\,} \sfL^- \to {\ff'}^- \, \cha_1^- \\
\phantom{\sfR^- \to \ff^- \, \neu_1, \quad \sfL^- \to {\ff'}^-\,}
  \knickpfeil \tau^- \nu_\tau \neu_1
\end{array}
\end{equation}
where $f'$ is the SU(2) doublet partner of $f$.
Fig.~\ref{fig:sp} shows the mass spectrum and branching ratios of the first and
second generation sleptons for the SPS1 scenario \cite{sps}.
%%%%%%%%%%%%%%%%%%%%%%%%%%%%%%%%%%%%%%%%%%%%%%%%%%%%%%%%%%%%%%%%%%%%
%%%%%%%%%%%%%%%%%%%%%%%%%%%%%%%%%%%%%%%%%%%%%%%%%%%%%%%%%%%%%%%%%%%%

\begin{figure}[tb]
(a)\\[1ex]
\epsfig{file=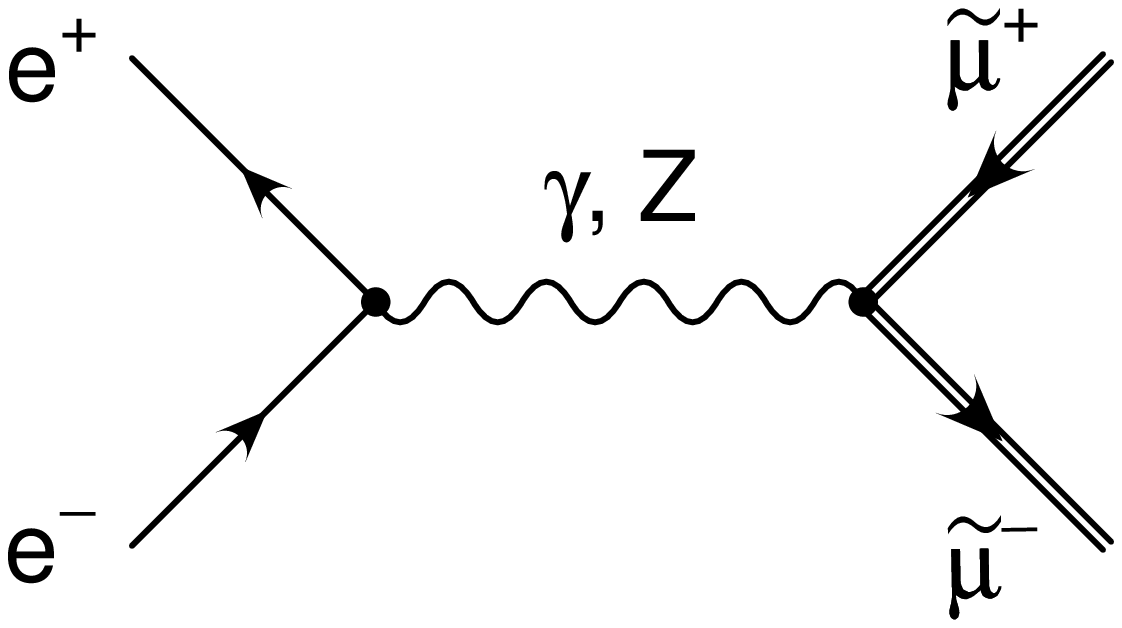,width=4cm}\\[1ex]
(b)\\
\epsfig{file=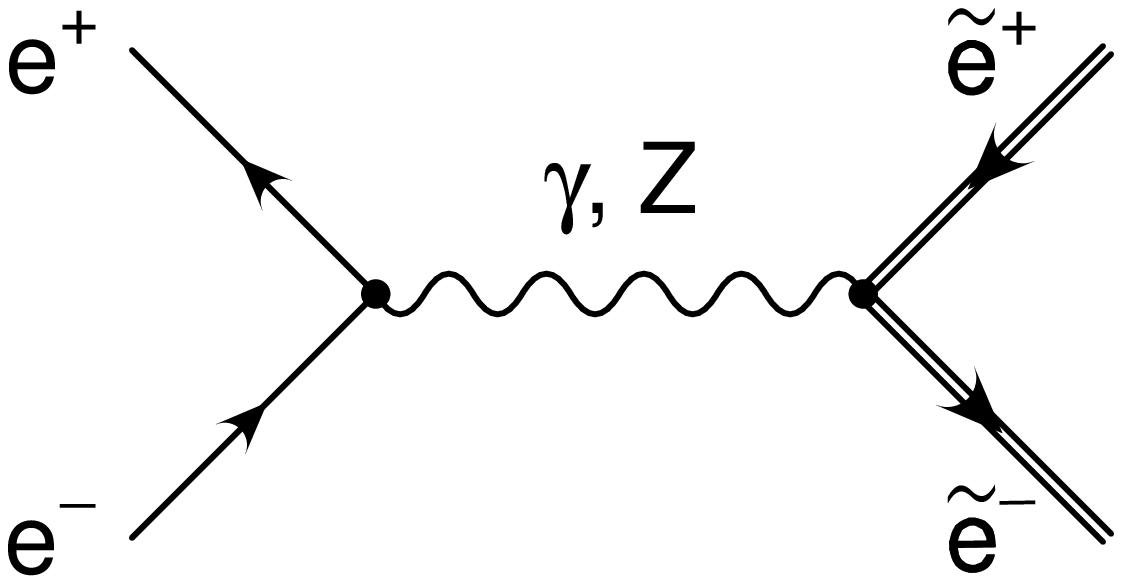,width=4cm}~\epsfig{file=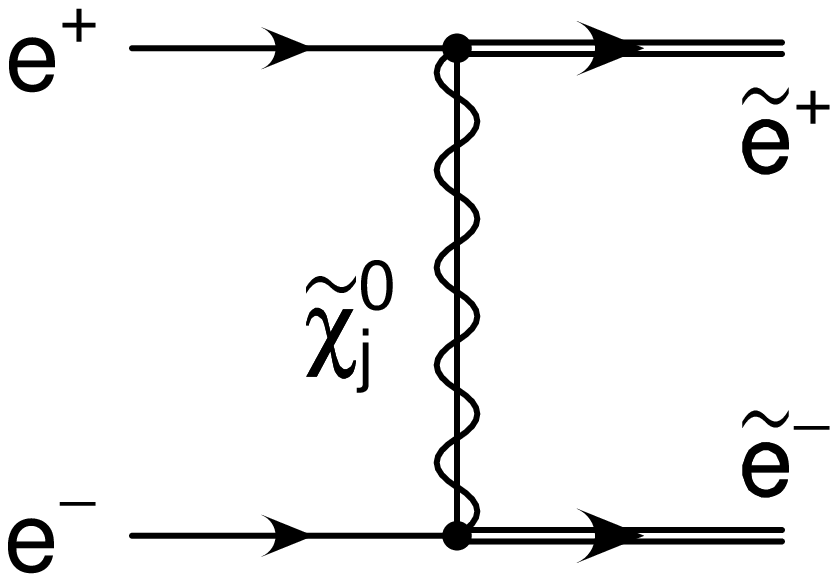,width=3.5cm}\\[1ex]
(c)\\[1ex]
\epsfig{file=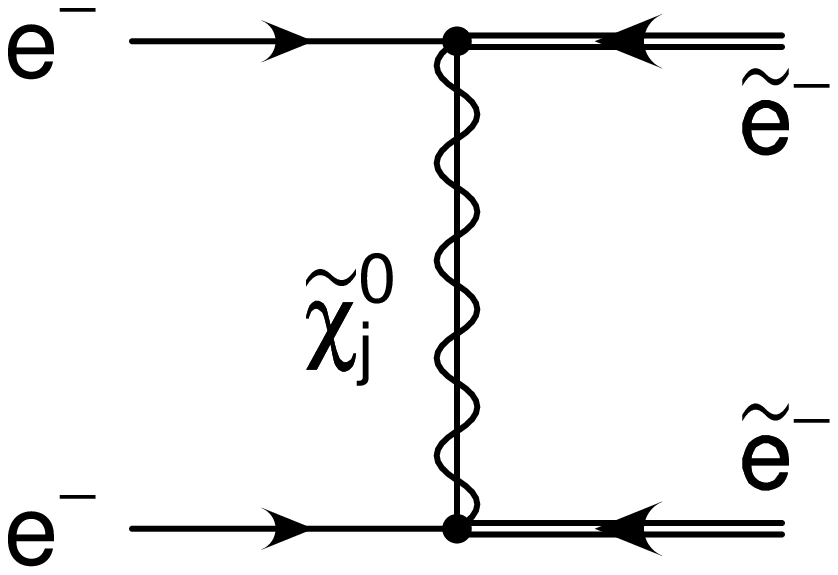,width=3.5cm}
\vspace{-2em}
\caption{Tree-level diagrams for the production of smuons (a) and selectrons (b)
in $e^+e^-$ annihilation and for the production of selectrons in $e^-e^-$
collisions (c).}
\label{fig:diag}
%\vspace{-1em}
\end{figure}

At leading order, sleptons of the second and third generation, as well
as squarks, are produced via s-channel gauge-boson exchange in $e^+e^-$
annihilation, see Fig.~\ref{fig:diag}~(a).%
%%%%%%%%%%%%%%%%%%%%%%%%%%%%%%%%%%%%%%%%%%%%%%%%%%%%%%%%%%%%%%%%%%%%
%%%%%%%%%%%%%%%%%%%%%%%%%%%%%%%%%%%%%%%%%%%%%%%%%%%%%%%%%%%%%%%%%%%%
As a consequence, they are produced in a P-wave with a characteristic rise of
the excitation curve $\sigma \propto \beta^3$, where
$\beta = \sqrt{1-4\msf^2/s}$ is the sfermion velocity.
Due to the Majorana nature of the neutralinos in the t-channel, see
Fig.~\ref{fig:diag}~(b) and (c), selectrons can
also be produced in S-waves ($\sigma \propto \beta$), namely for $\seR^\pm
\seL^\mp$ pairs in $e^+e^-$ annihilation and $\seR^-\seR^-, \seL^-\seL^-$ pairs
in $e^-e^-$ scattering.
The $e^-e^-$ mode has the additional advantage of much lower background.

\section{SELECTRON QUANTUM NUMBERS}

Due to the small masses of the first two generation fermions, the mixing
between the corresponding scalar partners is expected to be negligible. As a
consequence, the mass eigenstates $\sF_1$, $\sF_2$ are almost identical to the
chirality states $\sfL$, $\sfR$, which are the scalar partners of the left- and
right-handed states of the fermion $f$, respectively.

The additional t-channel amplitude for selectron production directly relates
the quantum numbers of the
produced selectrons and spositrons to the polarization of
the incoming electron and positron.
Therefore
the chiral quantum numbers of the selectrons can be unambiguously
determined by studying the selectron production cross-sections in $e^+e^-$
annihilation with polarized beams \cite{Bloechi:02}.
In order to separate the t-channel neutralino exchange
from the s-channel photon and Z-boson exchange,
both the electron and positron beams must be polarized. By comparing the
selectron cross-section for different polarization combinations the chiral
quantum numbers of the selectrons can be disentangled.

\section{SUSY COUPLING RELATIONS}

One
of the most fundamental predictions of supersymmetry is the equivalence
between gauge couplings and their supersymmetric Yukawa counterparts. For instance,
the gauge coupling between a vector boson $V$ and a
fermion $f$, $g(Vff)$,
is related to
the Yukawa coupling between the
gaugino partner $\tilde{V}$ of the vector boson, the fermion $f$ and the
sfermion $\tilde{f}$, $\hat{g}(\tilde{V}f\tilde{f})$. Within softly broken
supersymmetric theories %all three 
both kinds of couplings are required to be identical,
\begin{equation}
g =
\hat{g}.
\end{equation}

%%%%%%%%%%%%%%%%%%%%%%%%%%%%%%%%%%%%%%%%%%%%%%%%%%%%%%%%%%%%%%%%%%%%
\begin{table*}[bt]
\caption{Expected errors for the extraction of the supersymmetric
Yukawa couplings $\hat{g}'$ and $\hat{g}$, which correspond to the U(1)$_{\rm Y}$
and SU(2)$_{\rm L}$ gauge couplings $g'$ and $g$, respectively. No detector simulation
is included. The values are for the SPS1 scenario \cite{sps} and $\sqrt{s} = 500 \gev$.}
\label{tab:coupl}
\setlength{\tabcolsep}{1.5pc}
% -----------------------------------------------------
% adapted from TeX book, p. 241
\newlength{\digitwidth} \settowidth{\digitwidth}{\rm 0}
\catcode`?=\active \def?{\kern\digitwidth}
% -----------------------------------------------------
\begin{tabular*}{\textwidth}{@{}lrrr}
\hline
Process & Luminosity & \multicolumn{1}{c}{Result}\\
\hline
$e^+e^- \to (\seR^+ \seR^-)
	\to e^+ e^- \, \neu_1 \, \neu_1$ & 500 fb$^{-1}$ &
  $\delta \hat{g}'/\hat{g}' \approx 0.2 \%$ \\
$e^+e^- \to (\seR^\pm \seL^\mp)
	\to e^+ e^- \, \neu_1 \, \neu_2
	\to e^+ e^- \, \tau\tau \, \neu_1 \, \neu_1$ &
  & $\delta \hat{g}/\hat{g} \approx 0.7 \%$ \\
\hline
$e^-e^- \to (\seR^- \seR^-)
	\to e^- \, e^- \, \neu_1 \, \neu_1$ & 50 fb$^{-1}$ &
  $\delta \hat{g}'/\hat{g}' \approx 0.2 \%$ \\
$e^-e^- \to (\seL^- \seL^-(
	\to e^+ e^- \, \neu_2 \, \neu_2
	\to e^+ e^- \, 4\tau \, \neu_1 \, \neu_1$ &
 & $\delta \hat{g}/\hat{g} \approx 0.8 \%$ \\
\hline
\end{tabular*}
\end{table*}
%%%%%%%%%%%%%%%%%%%%%%%%%%%%%%%%%%%%%%%%%%%%%%%%%%%%%%%%%%%%%%%%%%%%
This relation can be tested in the electroweak sector by measuring the
production cross-sections of scalar leptons.
The Yukawa couplings $\hat{g}$
can best be probed in the production of selectrons, as a result of the t-channel
neutralino exchange contributions \cite{eYuk}. 
One can take advantage
of the $e^-e^-$ mode due to reduced background, larger cross-sections, higher
polarizability and no interfering s-channel contributions \cite{superoblique2}.
Simulations have shown that these couplings can be determined with high
accuracy by measuring the slepton cross-sections at a linear collider
\cite{mythesis,mysusy}, see Tab.~\ref{tab:coupl}. This method is complementary to the
measurement of the Yukawa couplings in chargino and neutralino pair production
\cite{CKMZ}.

Clearly the experimental precision requires the inclusion of radiative
corrections in the theoretical predictions for the slepton cross-sections. 
As a
first step the produced sleptons may be considered on-shell, since in the
continuum far above threshold the effect of the non-zero slepton width
is relatively small, of the order $\Gamma_{\!\sF}/\msf$.
As a consequence, the production and decay of the sleptons
can be treated separately. For both sub-processes, the complete electroweak
one-loop corrections in the MSSM have been computed \cite{slRC,Guasch:01}.
In
Ref.~\cite{slRC} the pair production of smuons in $e^+e^-$ and the production
of selectrons in $e^+e^-$ and $e^-e^-$ collisions have been calculated at
next-to-leading order.
The ${\cal O}(\alpha)$ corrections for the most important decay
channels of the sleptons, $\sll^\pm \to l^\pm \, \neu_i$ and
$\sll^\pm \to ^{^{_{(-)}}} \hspace{-3ex} \nu_l \, \cha^\pm_i$
have been considered in Ref.~\cite{Guasch:01}.

Both calculations have been performed in the on-shell renormalization scheme.
While the renormalization of the gauge couplings and gauge bosons can be
performed similar to the Standard Model case, the renormalization of the
superpartner contributions is slightly more involved. For example, the sector
of the charginos and neutralinos depends on the gaugino parameters  $M_1$,
$M_2$ and the Higgs parameters $\mu$ and $\tan\beta$.  It should be noted that
the definition of $\tan\beta$ cannot be related to any physical observable in a
straightforward manner \cite{stoecki:02}. Instead, for technical reasons, the
$\overline{\rm DR}$ scheme is used in \cite{slRC} for the renormalization of
$\tan\beta$. By demanding on-shell mass conditions for three out of the six
charginos and neutralinos, the parameters $M_1$, $M_2$ and $\mu$ can be fixed.
The other three mass eigenvalues then receive non-zero corrections at the
one-loop level. 
For more details on the renormalization see \cite{mythesis,slRC,Guasch:01}.

The effects of the electroweak corrections were found to be
sizeable, of the order of 5--10\%. They include important effects from
supersymmetric particles in the virtual corrections,
in particular non-decoupling logarithmic contributions,
e.g. terms $\propto \log m_{\sF}/m_{\rm weak}$ 
from fermion-sfermion-loops.

The equality of gauge and Yukawa couplings in the SU(3)$_{\rm C}$ gauge sector
can be tested at a linear collider by investigating the associated production
of quarks $q$ and squarks $\tilde{q}$ with a gluon $g$ or gluino $\tilde{g}$.
While the processes $e^+e^- \to q \bar{q} g$ and $e^+e^- \to \tilde{q}
\bar{\tilde{q}} g$ are sensitive to the strong gauge coupling of quarks and
squarks, respectively, the corresponding Yukawa coupling can be probed in
$e^+e^- \to q \bar{\tilde{q}} \tilde{g}$.
In order to obtain reliable
theoretical predictions for these cross-sections it is necessary to include
next-to-leading order (NLO) supersymmetric QCD corrections. These
corrections are generally expected to be rather large and they are necessary
to reduce the large scale dependence of the leading-order result.
The NLO QCD corrections to the process $e^+e^- \to q \bar{q} g$
within the Standard Model are known since long \cite{NLOqqg}.
Recently, the
complete ${\cal O}(\alpha_{\rm s})$ corrections to all three processes in the
MSSM have been calculated \cite{BMWZ}. 
The NLO contributions enhance the cross-section in the peak region by roughly
20\% with respect to the LO result. Furthermore, the scale dependence is
reduced by a factor of about six when the NLO corrections are included.

\section{MASS DETERMINATION}

In models with R-parity conservation, the mass of
supersymmetric particles cannot directly be reconstructed from its decay
products since the lightest supersymmetry particle (LSP)---mostly the lightest
neutralino $\neu_1$---escapes undetected. Nevertheless both
the mass of scalar fermions and the unobserved LSP can be determined
by measuring the end points of the kinematical energy distributions of the
visible decay fermions \cite{Tsukamoto:1995gt,martyn}.
However the efficiency
of this method is affected by initial-state radiation (ISR) and beamstrahlung
effects and cuts necessary to reduce the backgrounds.

\begin{figure}[tb]
\epsfig{figure=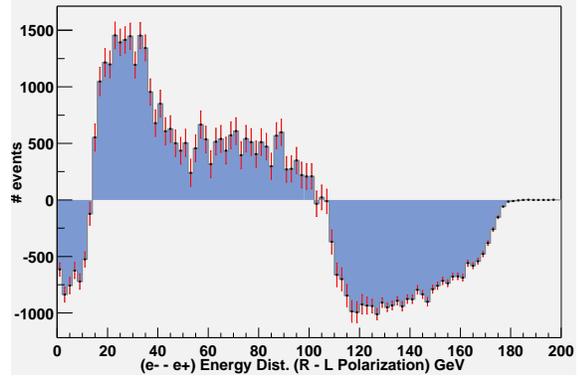, height=\columnwidth, angle=-90,
	viewport=40 20 480 680, clip=true}
\vspace{-2em}
\caption{Difference of the subtracted normalized electron and positron energy
distributions for 80\% L- and R-polarized $e^-$ beam.}
\label{fig:mdist}
%\vspace{-1em}
\end{figure}

A novel approach for determining the selectron masses \cite{Dima:01}
takes advantage of the fact that selectrons can be produced in mixed pairs,
$e^+e^- \to \seR^+ \seL^-$, $\seL^+ \seR^-$. Because of the mass difference
between L- and R-selectrons,
the minima and maxima of the final state lepton energy distributions originating
from an $\seR^+ \seL^-$ pair will be different for the final electron and
positron. When
subtracting the normalized electron and positron distributions, the relevant
kinematical edges will therefore remain visible while the Standard Model
background and a large part of supersymmetric backgrounds are canceled. 
In particular, the contributions from the production of selectrons with same
chirality, $\seR^+ \seR^-$ and $\seL^+ \seL^-$, are cancelled while the signal
from mixed pairs, $\seR^+ \seL^-$ and $\seL^+ \seR^-$ is preserved.
In order to disentangle $\seR^+ \seL^-$ pairs from $\seL^+ \seR^-$ pairs, 
appropriate polarization of the incident electron beam is used and
the subtracted
electron and positron distributions for left- and right-handed $e^-$ beam
polarization are again subtracted. The final distribution 
including ISR and beamstrahlung and
cascade decays shown in Fig.~\ref{fig:mdist} nicely
exhibits the characteristic kinematical
edges while being unaffected by backgrounds. 
With this method it is possible to determine the masses of the selectrons and
the neutralino LSP with a
precision of 200--500 MeV for an integrated luminosity of $\int L = 1000$
fb$^{-1}$.
%%%%%%%%%%%%%%%%%%%%%%%%%%%%%%%%%%%%%%%%%%%%%%%%%%%%%%%%%%%%%%%%%%%%
%%%%%%%%%%%%%%%%%%%%%%%%%%%%%%%%%%%%%%%%%%%%%%%%%%%%%%%%%%%%%%%%%%%%

Alternatively, the sfermion masses can be extracted from the measurement
of the pair production cross-sections near threshold. The characteristic
onset of the excitation curve allows a very precise determination of the
sfermion masses in threshold scans. Since the
expected experimental accuracy is of ${\cal O}(100 \mev)$, it is necessary to
incorporate effects beyond leading order in the theoretical predictions
\cite{thr1}. The non-zero widths of the sfermions, which considerably affect
the cross-sections near threshold, must be included in a gauge-invariant
manner. This can be achieved by shifting the sfermion mass into the
complex plane, $\msf^2 \to \msf^2 - i \msf \Gamma_{\!\sF}$.
Moreover, for the production of off-shell sfermions, 
the full $2\to4$ matrix element (e.g. $e^+e^- \to f^+
f^- \neu_1 \neu_1$)
must be taken into account, including the decay of the
sfermions as well as the MSSM backgrounds and interference contributions.
One of the most important radiative corrections near threshold is the Coulomb
rescattering correction due to photon exchange between the slowly moving
sfermions. For the production of off-shell sfermions with orbital angular
momentum $l$ it is given by
\begin{eqnarray}
%\hspace{-.5ex}
\sigma_{\rm coul} \hspace{-1.5ex}&=&\hspace{-1.5ex}  \sigma_{\rm born} \,
\frac{\alpha\pi}{2 \beta_p} Q^2_{\rm \sF} \nonumber \\[-.5ex]
&& \times
\biggl [ 1\! -\! \frac{2}{\pi} \arctan
\frac{|{\beta_M}|^2 - \beta_p^2}{2 \beta_p \;
        \Im \! m \, \beta_M} \biggr ] 
\Re e \,
        {\mathcal C}_l \!\!\; ,  \hspace{-1ex}\nonumber \\[.5ex]
%\hspace{-.5ex}
\quad {\mathcal C}_l \hspace{-1.5ex}&=&\hspace{-1.5ex}
  \left[\frac{\beta_p^2 + \beta_M^2}{2 \beta_p^2}
  \right]^l
\end{eqnarray}
with
\begin{eqnarray}
\beta_p \hspace{-1.5ex}&=&\hspace{-1.5ex} \frac{1}{s}\sqrt{(s-p_+^2-p_-^2)^2-4
p_+^2 p_-^2}, \\
\beta_M \hspace{-1.5ex}&=&\hspace{-1.5ex} \frac{1}{s}\sqrt{(s-M_+^2-M_-^2)^2-4
M_+^2 M_-^2}.
\end{eqnarray}
$Q_{\rm \sF}$, $p_\pm$ and $M^2_\pm = m_\pm^2 - i \, m_\pm \Gamma_{\!\pm}$ are
the charge quantum number, the momenta and complex pole masses of the
produced sfermion and anti-sfermion.
Beamstrahlung and ISR also play an important role.
As mentioned in sec.~\ref{sec:phen}, selectrons can in addition be produced in
$e^-e^-$ collisions.

%%%%%%%%%%%%%%%%%%%%%%%%%%%%%%%%%%%%%%%%%%%%%%%%%%%%%%%%%%%%%%%%%%%%
\begin{figure}[tb]
 $e^+e^- \to (\seR^+\seR^-) \to e^+e^- + \Eslash$ \\[.5ex]
\epsfig{file=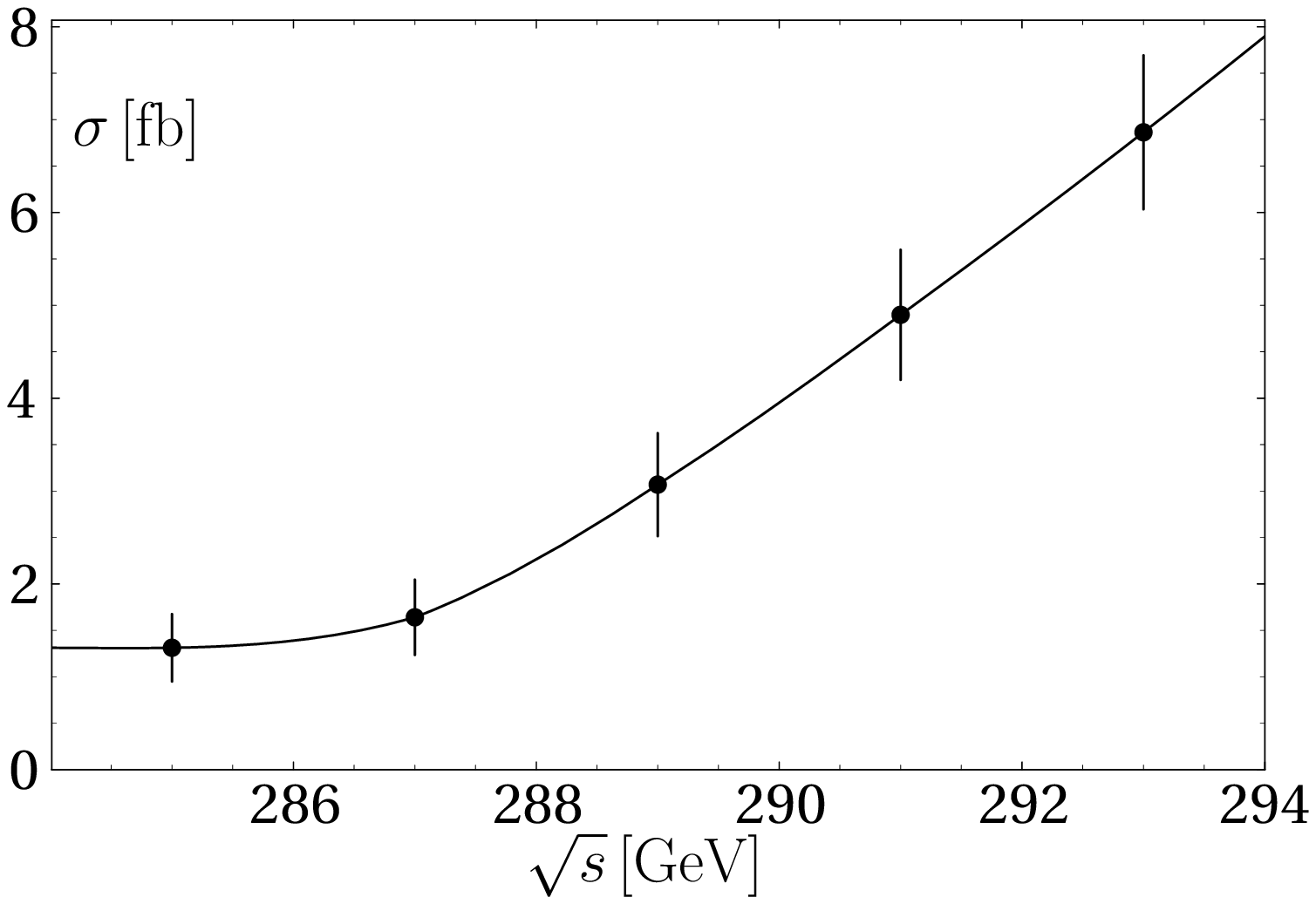,width=\columnwidth,
	viewport=30 0 472 290, clip=false} \\[1ex]
 $e^-e^- \to (\seR^-\seR^-) \to e^-e^- + \Eslash$ \\[.5ex]
%\rule{0mm}{0mm}
\epsfig{file=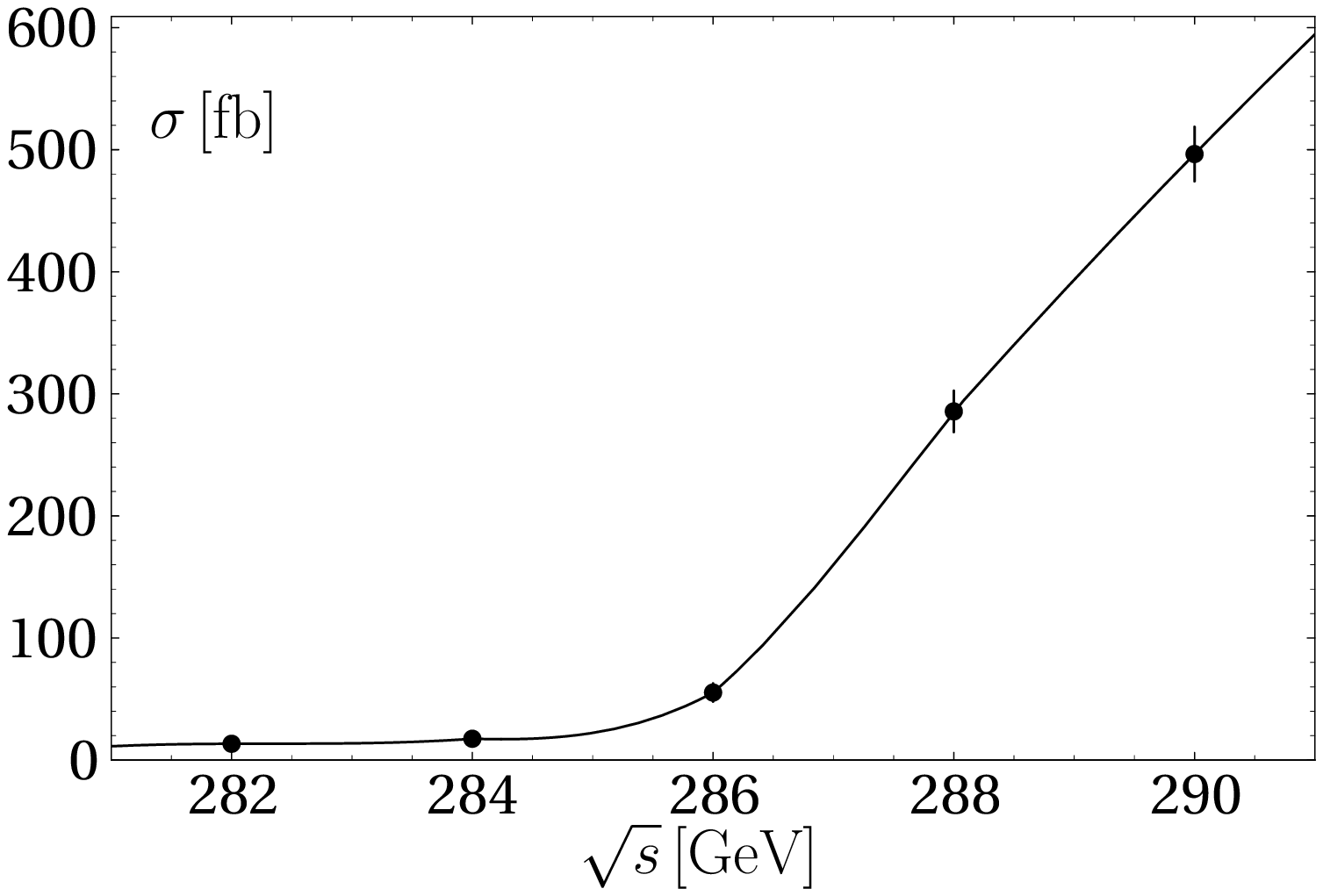,width=\columnwidth,
	viewport=43 0 472 290, clip=false}
\vspace{-3em}
\caption{Threshold excitation curves for R-selectron pair production. The error
bars indicate the statistical error for a luminosity per scan point of $\int L=10
{\rm \ fb}^{-1}$ in $e^+e^-$ and $\int L=1
{\rm \ fb}^{-1}$ in $e^-e^-$ mode, respectively (SPS1 scenario).}
\label{fig:thrsel}
%\vspace{-1em}
\end{figure}
%%%%%%%%%%%%%%%%%%%%%%%%%%%%%%%%%%%%%%%%%%%%%%%%%%%%%%%%%%%%%%%%%%%%
\begin{figure}[tb]
 $e^+e^- \to (\smuR^+\smuR^-) \to \mu^+\mu^- + \Eslash$ \\[.5ex]
\epsfig{file=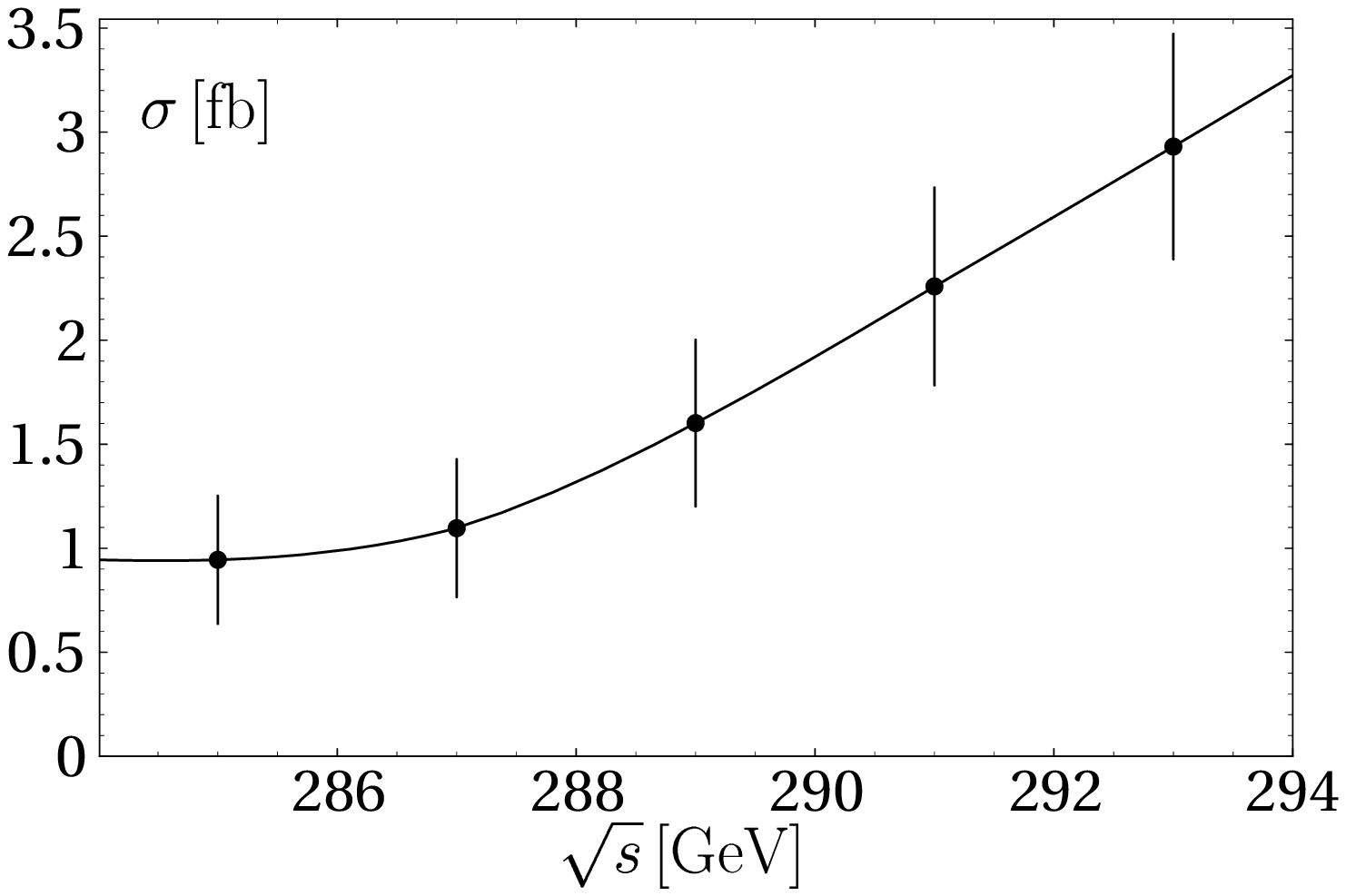,width=\columnwidth,
	viewport=43 0 472 290, clip=false}
\vspace{-3em}
\caption{Threshold excitation curve for R-smuon pair production with %statistical
error bars corresponding to $\int L=10 {\rm \ fb}^{-1}$ per scan point (SPS1
scenario).}
\label{fig:thrsmu}
%\vspace{-1em}
\end{figure}
%%%%%%%%%%%%%%%%%%%%%%%%%%%%%%%%%%%%%%%%%%%%%%%%%%%%%%%%%%%%%%%%%%%%
As an example,
the selectron cross-sections for both collider modes 
and the cross-section for smuon production in $e^+e^-$ annihilation
are shown in
Fig.~\ref{fig:thrsel} and Fig.~\ref{fig:thrsmu},
including five equidistant scan points.
Using four free parameters, the %sfermion
mass, width, normalization and a flat
background contribution, it is possible to fit the excitation cross-section
in a model-independent way. Results for the production of
selectrons %and smuons
are given in Tab.~\ref{tab:thrmass}. The numbers have been
obtained
with a MC generator without detector simulation but
including ISR and the theoretical refinements mentioned above.
The backgrounds from both SM and
MSSM sources, reduced by appropriate cuts, have also been included.
%%%%%%%%%%%%%%%%%%%%%%%%%%%%%%%%%%%%%%%%%%%%%%%%%%%%%%%%%%%%%%%%%%%%
\begin{table}[bt]
\caption{Expected precision for the determination of selectron masses and widths
from threshold scans \cite{mythesis,mysusy}. The values are for the
SPS1 scenario.}
\label{tab:thrmass}
\setlength{\tabcolsep}{2mm}
\renewcommand{\arraystretch}{1.3}
\begin{tabular*}{\columnwidth}{@{}p{2.7cm}l@{$\;$}r@{$\;\;\;$}r@{}}
\hline
Process & Lumn. & \multicolumn{1}{c}{Mass} & Width \\[-1ex]
&  [fb$^{-1}$] &  \multicolumn{1}{c}{[GeV]} & [MeV] \\
\hline
 & & \multicolumn{1}{c}{$\mseR$} & \multicolumn{1}{c}{$\;\;\Gamma_{\!\seR}$} \\
$e^+e^- \!\!\to (\seR^+ \seR^-) \newline
\phantom{e^+e^- } \!\!\to e^+ e^- + \Eslash$ & $5\!\times\! 10$ & $143.0^{+0.21}_{-0.19}$ &
  $150^{+300}_{-250}$ \\
$e^-e^- \!\!\to (\seR^- \seR^-) \newline
\phantom{e^-e^- } \!\!\to e^- e^- + \Eslash$ & $5\!\times\! 1$ & $142.95^{+0.048}_{-0.053}$ &
  $200^{+50}_{-40}$ \\
\hline
 & & \multicolumn{1}{c}{$\mseL$} & \multicolumn{1}{c}{$\;\;\Gamma_{\!\seL}$} \\
$e^+e^- \!\!\to (\seR^\pm \seL^\pm)$ &
  $\;\;5\!\times\! 10$ & $202.2^{+0.37}_{-0.33}$ &
  $240^{+40}_{-40}$ \\[-.5ex]
\multicolumn{2}{@{}l}{$\phantom{e^+e^- } \!\!\to e^+ e^- \tau\tau +\Eslash$} \\
$e^-e^- \!\!\to (\seL^- \seL^-)$ &
  $\;\;5\!\times\! 1$ & $202.2^{+0.62}_{-0.44}$ &
  $240^{+500}_{-400}$ \\[-.5ex]
\multicolumn{2}{@{}l}{$\phantom{e^-e^- } \!\!\to e^- e^- 4\tau +\Eslash$} \\
\hline
\end{tabular*}
\end{table}
%%%%%%%%%%%%%%%%%%%%%%%%%%%%%%%%%%%%%%%%%%%%%%%%%%%%%%%%%%%%%%%%%%%%

These high-precision measurements are necessary for reconstructing the
soft-breaking parameters at the GUT and Planck scales \cite{BPZ:00}.
As an example,
in \cite{Baer:01} a method is proposed to probe a non-universality between the
left and right slepton masses at the GUT scale if only both selectrons
and the light chargino are accessible at a linear collider. Inspired
by the one-loop RGE for slepton masses in mSUGRA-type models the
quantity
\begin{equation}
\begin{array}{@{}r@{\,}l@{}}
\Delta = &
m_{\tilde{e}_R}^2 - m_{\tilde{e}_L}^2 
+ {{m_{\cha^\pm_1}^2}\over {2\alpha_2^2(m_{\cha^\pm_1})}}
\bigl[{3\over 11}\bigl(\alpha_1^2(m_{\tilde{e}})\\[2ex]
 & -\alpha_1^2(M_{\rm GUT})\bigr)
-3\bigl(\alpha_2^2(m_{\tilde{e}}) -\alpha_2^2(M_{\rm GUT})\bigr)\bigr]
\end{array}
\end{equation}
is defined which is strongly correlated with the slepton mass
non-universality
$
\delta m^2 \equiv m_{\tilde{e}_R}^2(M_{\rm GUT}) - m_{\tilde{e}_L}^2(M_{\rm GUT})
$.
The method allows to detect a non-universality if $|\delta m^2| \gesim
5\cdot 10^3 \gev^2$. Using additional information on $M_2$ this
improves to $|\delta m^2| \gesim 2\cdot 10^3 \gev^2$, i.e. $\delta m/m \gesim 1.6\%$
for $m \approx 250 \gev$.

\section{SINGLE SLEPTON PRODUCTION}

In general, only sfermions with masses up to the beam energy, $\msf <
\sqrt{s}/2$, can be investigated using the pair production process. Below the pair
production threshold, sfermions can only be produced in association with their
partner fermions and charginos or neutralinos. With sufficiently high luminosity
at a future linear collider ($\int L=500 {\rm \ fb}^{-1}$), it is possible to probe
sfermions with masses that exceed the beam energy by a few tens of GeV
if the gauginos are relatively light \cite{datta:0102}.
Due to the additional photon exchange in the
t-channel, the discovery potential
of the associated production is particularly promising for selectrons.

\section{THIRD GENERATION SFERMIONS}

In contrast to the first two generation sfermions, large mixings are expected
between the left- and right-chiral states of the third generation sfermions
due to the large Yukawa coupling.
It can be seen from the sfermion mass matrix
that for large fermion masses $m_{\rm f}$ the sensitivity to the Higgs parameters
$\mu$ and $\tan\beta$ as well as the trilinear scalar couplings $A_{\rm f}$ is
enhanced:
\begin{equation}
M_{\rm \tilde{f}}^2 = \!\left( \begin{array}{@{}c@{\;\;\;}c@{}}
m_{\rm f}^2 + m_{\rm \tilde{F}_L}^2 + D_{\rm L} &
 m_{\rm f} \bigl( A_{\rm f} - \mu/t_\beta^{2 I^3_{\rm f}} \bigr) \\
m_{\rm f} \bigl( A_{\rm f} - \mu/t_\beta^{2 I^3_{\rm f}} \bigr) &
m_{\rm f}^2 + m_{\rm \tilde{f}_R}^2 + D_{\rm R}
\end{array} \right)\!\!,
\end{equation}
where $m_{\rm \tilde{F}_L}$, $m_{\rm \tilde{f}_R}$, $D_{\rm L}$, $D_{\rm R}$
are the soft-breaking masses and D-terms of the sfermions $\tilde{f}_{\rm L,R}$,
and $I^3_{\rm f}$ is the weak isospin quantum
number of the fermion $f$. The shortcut $t_\beta \equiv \tan\beta$ has been
used.

\begin{figure}[tb]
\epsfig{figure=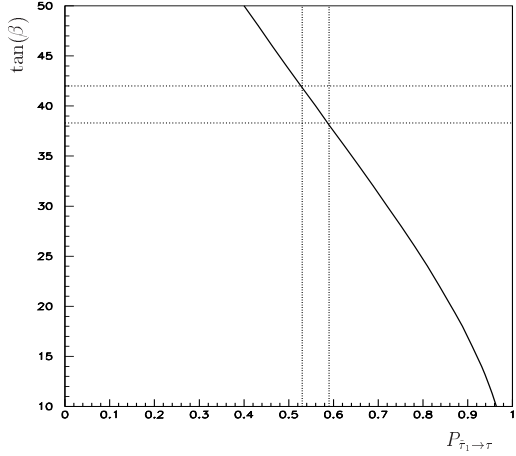, width=\columnwidth,
	viewport=0 50 402 335}
%\epsfig{figure=tanxi10.epsi, width=\columnwidth,
%	viewport=30 80 526 547, clip=true}
%\vspace{-2em}
\caption{Determination of $\tan\beta$ from the tau polarization 
$P_{\tilde{\tau}_1 \to \tau}$ in the decay $\tilde{\tau}^-_1 \to \tau^- \neu_1$.
}
\label{fig:taupol}
%\vspace{-1em}
\end{figure}

For instance, the production and decay of staus is interesting for the
determination of $\tan\beta$, in particular at high values\footnote{In this
region the chargino/neutralino analyses are not effective~\cite{CKMZ}.},
by examining the
polarization of the taus in the decay $\tilde{\tau}^-_{1,2} \to \tau^-_{\rm
L,R} \neu_1$ \cite{Nojiri,eYuk}. In a recent study within the MSSM
\cite{Boos:02} it was found that $\tan\beta$ in the range of 30--40 can be
determined with an error of about 10\% when the $\tau$
polarization is measured only for $\tilde{\tau}_1$ decays
(see also Fig.~\ref{fig:taupol}),
assuming a sufficiently large higgsino component of the lightest neutralino
$\neu_1$.
%%%%%%%%%%%%%%%%%%%%%%%%%%%%%%%%%%%%%%%%%%%%%%%%%%%%%%%%%%%%%%%%%%%%
%%%%%%%%%%%%%%%%%%%%%%%%%%%%%%%%%%%%%%%%%%%%%%%%%%%%%%%%%%%%%%%%%%%%

The parameters $\mu$ and $A_\tau$ as well as the neutralino parameter $M_1$ may
carry CP-violating phases. By using measurements
of stau masses, cross-sections and branching ratios, it is possible to
extract these parameters including their phases from the stau system.
Very high accuracy on all involved
parameters
can be obtained \cite{Bartl:02} if the decay of the heavy staus into heavy Higgs
bosons is kinematically allowed%
, see Tab.~\ref{tab:CPtau}.
%%%%%%%%%%%%%%%%%%%%%%%%%%%%%%%%%%%%%%%%%%%%%%%%%%%%%%%%%%%%%%%%%%%%
\begin{table}[bt]
\caption{Extracted parameters from simulated data of the masses, cross-sections
and BRs of $\tilde{\tau}_i$ for
$\sqrt{s} = 800 \gev$ and $\int L = 500$ fb$^{-1}$. Original values:
$M_{\tilde{\tau}_{\rm L}} = 350 \gev$, $m_{\tilde{\tau}_{\rm R}} = 150 \gev$,
$A_\tau = -800 i \gev$, $M_2 = 280 \gev$, $\mu = 250 \gev$, $\phi_{\rm U(1)} =
0$.}
\label{tab:CPtau}
\setlength{\tabcolsep}{.5pc}
\renewcommand{\arraystretch}{1.3}
\begin{tabular*}{\columnwidth}{@{}l@{$\;\;$}cc}
\hline
$\tan\beta$ & 3 & 30 \\
\hline
$m_{\tilde{\tau}_{\rm R}}^2$ [GeV$^2$] &
  $(225 \pm 2) \cdot 10^2$ & $(225 \pm 6) \cdot 10^2$ \\
$m_{\tilde{\tau}_{\rm L}}^2$ [GeV$^2$] &
  $(1225 \pm 4) \cdot 10^2$ & $(1229 \pm 7) \cdot 10^2$ \\
$\re  A_\tau$ [GeV] &
  $-8 \pm 180$ & $8 \pm 55$ \\
$\im  A_\tau$ [GeV] &
  $-800 \pm 70$ & $-800 \pm 21$ \\
\hline
$\re \mu$ [GeV] &
  $249.9 \pm 0.26$ & $249.9 \pm 0.6$ \\
$\im \mu$ [GeV] &
  $2.4 \pm 1.7$ & $-0.2 \pm 3.8$ \\
$\tan\beta$ & $2.99 \pm 0.03$ & $29.9 \pm 0.7$ \\
\hline
$\re M_1$ [GeV] & $140.9 \pm 0.2$ & $140.6 \pm 0.6$ \\
$\im M_1$ [GeV] & $-0.7 \pm 3.4$ & $0.16 \pm 1.0$ \\
$M_2$ [GeV] & $280.0 \pm 0.3$ & $280 \pm 1$ \\
\hline
\end{tabular*}
\end{table}
%%%%%%%%%%%%%%%%%%%%%%%%%%%%%%%%%%%%%%%%%%%%%%%%%%%%%%%%%%%%%%%%%%%%

CP-violating parameters in the stau sector can also give rise to
electric and weak dipole moments of the taus. At a linear collider with high
luminosity and polarization of both $e^+$ and $e^-$ beams it is possible to
probe CP-violating tau dipole form factors up to the level of ${\cal O}(3)-
{\cal O}(5) \cdot 10^{-19} e {\rm cm}$ \cite{Anan:02}. While this precision
would improve the current experimental bounds by three orders of magnitude, it
is however still about one order of magnitude beyond the
expectations from supersymmetric models with CP-violation.

The study of scalar top quarks is of particular interest since due to large
mixing the lighter stop mass eigenstate is likely to be the lightest squark. As
a consequence, the decay channel into $t\neu_1$ may be kinematically forbidden
so that other dedicated search strategies are necessary.  The most promising
decay modes in this case are $\tilde{t}_1 \to c \neu_1$ and $\tilde{t}_1 \to b
\cha_1^\pm$ \cite{Nowak:00}. A simulation including full-statistics
SM backgrounds
and a detailed investigation of quark tagging and
supersymmetric backgrounds
for the SPS5 scenario\footnote{The SPS5 scenario is a dedicated "light-stop"
scenario} \cite{sps} has been performed \cite{Nowak:02}.
Using the decay
channel $\tilde{t}_1 \to c \neu_1$,
the mass of the lighter stop and the mixing
angle can be constrained up to $\Delta m_{\rm \tilde{t}_1} \approx 1 \gev$ and
$\Delta \cos \theta_{\rm \tilde{t}_1} \approx 0.0125$
with a luminosity of 1000 fb$^{-1}$%
, while other decay
channels like $\tilde{t}_1 \to b \cha_1^\pm$ and $\tilde{t}_1 \to b \, l \,
\tilde{\nu}_l$ are not open in this specific scenario.

For stops and sbottoms with large mass differences, bosonic decay modes, $\tilde{t}_1 \to
\tilde{b}_1 + (H^+$ or $W^+)$, $\tilde{b}_1 \to
\tilde{t}_1 + (H^-$ or $W^-)$, may become
important.
Due to the large Yukawa
couplings and mixings of $\tilde{t}$ and $\tilde{b}$ these decay modes can
dominate in a wide range of MSSM parameters \cite{Bartl:01}.
Furthermore the bosonic decay modes are affected by large radiative SUSY-QCD
corrections \cite{Bartl:99}, whose leading contribution can be
taken into account by SUSY-QCD running parameters for the
stop/sbottom masses and the trilinear couplings. As a consequence, the precise
investigation of these decay modes is important for stop/sbottom searches and the
determination of the underlying MSSM parameters%
 at a future linear collider.

\vspace{1em}
\noindent {\bf Acknowledgements:} This work was supported in part by the
  European Commision 5th framework under contract HPRN-CT-2000-00149 and
  the Polish-German LC project No. POL 00/015. W.~P.~is
  supported by the "Erwin Schr\"odinger fellowship No.~J2095" of
the "Fonds zur F\"orderung der wissenschaftlichen Forschung" of Austria
FWF and partly by the Swiss "Nationalfonds".

\end{document}